\documentstyle[psfig,twoside,fleqn,espcrc2]{article}
\title{Grand-Canonical simulation of 4D simplicial quantum gravity
\thanks{presented by S.Horata}}

\author{S.Horata\address{Department of
Particle and Nuclear Physics, School of Mathematical Science,
The Graduate University for Advanced Studies,
Tsukuba, Ibaraki 305-0801, Japan}
,
H.S.Egawa\address{Department of Physics, Tokai University,
        Hiratsuka, Kanagawa 259-1292, Japan}
and
        T.Yukawa\address{Coordination Center for Research and Education,
        The Graduate University for Advanced Studies,
        Hayama, Miura, Kanagawa 240-0193, Japan}
}
\begin{document}
\begin{abstract}
A thorough numerical
examination for the field theory of 4D quantum
gravity (QG) with a special emphasis on the conformal mode dependence has
been studied.
More clearly than before, we obtain the string susceptibility
exponent of the partition function by using the Grand-Canonical
Monte-Carlo method.
Taking thorough care of the update method, the simulation is made for 4D
Euclidean
simplicial manifold coupled to $N_X$ scalar fields and $N_A$ U(1) gauge
fields.
The numerical results suggest that 4D simplicial quantum gravity (SQG)
can be reached to the continuum theory of 4D QG.
We discuss the significant property of 4D SQG.
\end{abstract}
\maketitle
\section{Introduction}
Until now 4D SQG has
been investigated from several points of view, such as the phase structure,
nature of the transition, effects of additional matter
fields, and the modified measure action \cite{T}.
Unfortunately the numerical results could not resolve the problem for the
existence of the continuum theory of 4D QG clearly.
Gradually the original enthusiasm to explore 4D QG has been toned down.
However, we remind that many problems about 4D SQG are yet to be studied.

As one of those problems, we pay attention to additional matter
effects in 4D Euclidean SQG (Eucl.SQG).
According to the recent numerical simulations, the string
susceptibility exponent ($\gamma^{(4)}$) takes a negative value with
adding a few ${\rm U(1)}$ gauge fields.
It makes drastic change of the phase diagram, and continuous phase
transition with one gauge matter has
been found\cite{PISA}.
Moreover, the numerically obtained $\gamma^{(4)}$ as a function of the number
of matter fields turned out to be approximately equal to that of the
prediction of 4D conformal field
theory\cite{BERLIN}.

In order to see more definitely how the matter fields act on 4D
Eucl.SQG, we repeat the study with employing the Grand-Canonical method used
for 2D SQG\cite{OTY}, which has given the satisfactory agreement of two
results.
For discussing the continuum theory of 4D QG based on the numerical
analysis, we pay attention to the functional form of the partition
function, which we expect enough to gain information for selecting the
correct
theory of gravity, if it exists at all.
For obtaining the functional form of the partition function,
we need to perform a simulation with varying the space volume.
In practice, we calculate the probability ($P(N_4)$) as a function of the
volume ($N_4$) by the Grand-Canonical method of the dynamical
triangulation (DT) and estimate $\gamma^{(4)}$ from the scaling
behavior of $P(N_4)$.
%
\section{Model and
Grand-Canonical method}
Let us explain our model and the numerical method of the Grand-Canonical
simulation.
We consider a system of the simplicial gravity coupled to ${\rm U(1)}$
gauge fields and massless scalar fields.
The action ($S$) is given on 4D simplicial manifold as
\begin{equation}
 S = S_{G} + S_{A} + S_{X}, \label{eq.TotalAction}
\end{equation}
where $S_{G}$, $S_{A}$ and $S_{X}$ denote the action for the gravity, ${\rm
U(1)}$ gauge fields ($A$) and scalar fields ($X$),
respectively.

For the gravity part, we use the discretized Einstein-Hilbert action in 4D,
\begin{equation}
 S_{G}[\kappa_{2},\kappa_{4}]=
\kappa_{4}N_{4}-\kappa_{2}N_{2},
\end{equation}
where $N_{i}$ denotes the number of $i$-simplex.
The two parameters, $\kappa_2$ and $\kappa_4$, correspond to the inverse
of the gravitational constant and the cosmological constant,
respectively.
The action for the $N_X$ scalar fields $X_i^a$, $a=1...N_X$,
on the vertex $i$ is given by
\begin{equation}
 S_X = \sum_{a=1}^{N_X} \sum_{ij} o(l_{ij}) (X_i^a - X_j^a)^2,
\end{equation}
where $o(l_{ij})$ is the number of four-simplices sharing a link $l_{ij}$.
And the action for the $N_A$ vector field $A_{ij}^a$( $a=1...N_A$) on link
$l_{ij}$ reads
\begin{equation}
  S_A = \sum_{a=1}^{N_A} \sum_{ijk} o(t_{ijk})
(A^a(l_{ij})+A^a(l_{jk})+A^a(l_{ki}))^2,
\end{equation}
where $o(t_{ijk})$ is the number of four-simplices sharing a triangle
$t_{ijk}$.
These actions give the correct thermo-dynamical limit for large $N_4$
because they are proportional to $N_4$.

In order to simplify the calculation,
the Monte-Carlo simulations have been performed with the Canonical
method previously.
This method requires that $N_4$ fluctuates around the
given target volume ($V_4$).
Thus, the volume fixing term ($\delta S$) is added to the total action
Eq.(\ref{eq.TotalAction}),
\begin{equation}
 \delta S(N_4, V_4) = \delta \kappa_4 \left( N_4 - V_4 \right)^2, 
\end{equation}
where $\delta \kappa_4$ is the free parameter.
However, the Canonical method has limitations about the finite size
effect and the ergodic property.
Furthermore, the functional form of the size dependence of the partition
function
cannot be given clearly.
For the improvement of numerical accuracy we perform the simulations varying
the space volume by the Grand-Canonical method.
 The Grand-Canonical partition function with the matter fields is given as
\begin{eqnarray}
 Z(\kappa_{2},\kappa_{4},N_A,N_X) = \sum_{T}
  e^{-S_{G}(\kappa_{2},\kappa_{4})} \qquad\qquad\quad& & \nonumber \\
\prod_{N_A,N_X}
  \left(
   \int \prod_{l \in T} dA_l e^{-S_{A}(A_l)}
   \int \prod_{i \in T} dX_i e^{-S_{X}(X_i)}
 \right),
\end{eqnarray}
where all geometries $T$ and matter fields $A,X$ are updated using the
Metropolis method keeping the topology of discretized surface to be
$S^{4}$.

From the analogy of the 2D case , we assume that the partition function
obeys the KPZ like formula for the large-$N_4$ limit as
\begin{equation}
 Z \approx \sum_{N_4} N_4^{\gamma^{(4)}(\kappa_2)-3} e^{\left(\kappa_4 -
\kappa_4^c\left(\kappa_2\right)\right) N_4}.
\end{equation}
Then the probability $P(N_4)$ for the large-$N_4$ behaves approximately as
\begin{equation}
 P(N_4) \propto N_4^{\gamma^{(4)} -3}e^{-(\kappa_4-\kappa_4^c)N_4}.
\end{equation}
By tuning $\kappa_4$ close to $\kappa_4^c$, the
exponential growth is suppressed and $P(N_4)$ is given as the power
function of $N_4$.
The parameter $\kappa_4$ is determined in the following manner:
If $\kappa_4$ is chosen to be less (or greater) than $\kappa_4^c$,
the system will expand (or shrink) exponentially
in the Grand-Canonical simulation.
By requiring an exponential increase (or decrease)
to disappear in the simulation, we can find $\kappa_4$ precisely with
the following form,
\begin{equation}
 \frac{\ln P(N_4)}{\ln N_4} = (\gamma^{(4)} -3) -
(\kappa_4-\kappa_4^c)\frac{N_4}{\ln N_4} + O(1/\ln N_4).
\end{equation}
After tuning $\kappa_4$, we generate the configurations constraining
$N_4$ within an upper- and a lower-bound
($N_4^{min} \leq N_4 \leq N_4^{max}$).
%
\section{Numerical result}
In this section, we report numerical results with the Grand-Canonical
method.
\begin{figure}
\centerline{\psfig{file=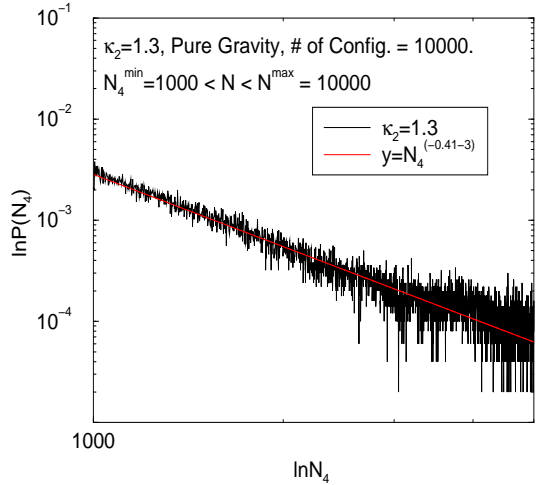,height=6.5cm,width=7cm}}
\vspace{-12mm}
\caption
{
 We plot $P(N_4)$ versus $N_4$ with log-log scale.
}
\vspace{-0.7cm}
\label{fig.lnP}
\end{figure}
In order to obtain $P(N_4)$, we allow $N_4$ to fluctuate between $N_4^{min}
= 1000$ and $N_4^{max} = 10000$.
In Fig.\ref{fig.lnP}, we show $P(N_4)$ versus $N_4$ at $\kappa_4$
close to $\kappa_4^c$.
The numerical result shows that $P(N_4)$ is given as a power
function of $N_4$.
It is suggested that the partition function in 4D obeys the KPZ
type partition function as the 2D case.

\begin{figure}
\centerline{\psfig{file=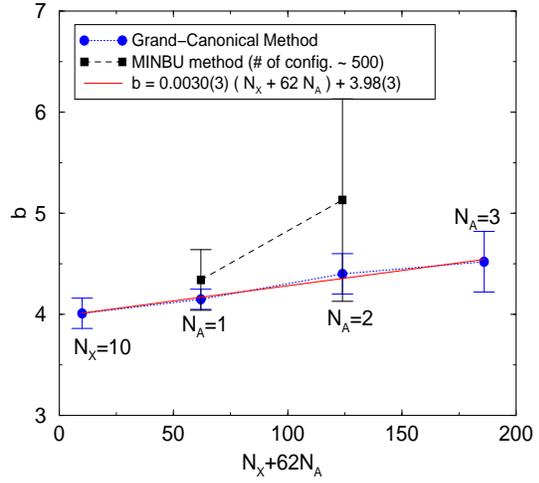,height=6.5cm,width=7cm}}
\vspace{-12mm}
\caption
{
We plot $b$ versus the number of the matter fields, $N_X + 62 N_A$.
}
\vspace{-0.7cm}
\label{fig.b}
\end{figure}

We can also compute $\gamma^{(4)}$ from the slope of $P(N_4)$.
Then, we show the parameter ($b$), which is calculated from
$\gamma^{4}$,
\begin{equation}
 \gamma^{(4)} = - \frac{b}{2} \left( 1 - \frac{4}{b} + \sqrt{ 1 - \frac{4}{b}}\right).
\label{eq.Gamma}
\end{equation}
This relation is given from the analogy of 2D QG
and the parameter $b$ is similar to the central charge in 2D QG.
%
%
%
In Fig.\ref{fig.b}, we plot $b$ versus the number of matter fields,
$N_X+62N_A$, on the critical point ($\kappa_2^c$) between the crumpled
phase and the smooth phase.
We compare $b$ obtained by two different numerical methods, {\it i.e.} the
Canonical method with the MINBU algorithm and the Grand-Canonical method.
Error bars include both the errors in determining $\kappa_2^c$ and
the statistical errors, and we found that the former
method has much larger errors than the latter method.
We made the parameter fit of the data of the Grand-Canonical method by
a linear function, $b = 0.0030(3) \left( N_X + 62 N_A \right) + 3.98(3)$.
The slope $b$ is approximately equal to the analytical
result\cite{PTP-LETTER,HAMA}.
\section{Summary and Discussion}
Let us summarize and discuss the main points made in the previous
section.
From the numerical results of the Grand-Canonical method,
we find the KPZ type partition function also appears in 4D QG.
The string susceptibility exponent $\gamma^{(4)}$ obtained from the size
dependence of the partition function can be obtained by the Grand-Canonical
method rather accurately.
The matter contribution to $b$, which is computed from $\gamma^{(4)}$
with the analogy of 2D QG, turns out to be approximately equal to the
conformal part\cite{PTP-LETTER,HAMA}.
Our numerical results suggest that it may be possible to understand the
continuum theory of 4D QG from the analogy of 2D QG.
In order to look more details on the relation to the continuum theory of 4D
QG,
these results encourage us to continue investigation by the Grand-Canonical
method and push more efforts on the analytical calculation of the conformal
gravity at the same time.


\begin{thebibliography}{99}
\bibitem{T}
G. Thorleifsson, Nucl. Phys. B (Proc. Suppl.) {\bf 73} (1998) 133.
\bibitem{PISA}
H. S. Egawa, S. Horata, N. Tsuda and T. Yukawa, Prog. Theor. Phys. 106
        (2001) 1037.
\bibitem{BERLIN}
H. S. Egawa, S. Horata, and T. Yukawa, Nucl. Phys. B
       (Proc. Suppl.) {\bf 106-107} (2002), 971.
\bibitem{OTY}
S. Oda, N. Tsuda and T. Yukawa, Nucl. Phys. B (Proc. Suppl.) {\bf 63}
 (1998) 733
\bibitem{PTP-LETTER}
H. S. Egawa, S. Horata, and T. Yukawa, Prog. Theor. Phys. submitted.
\bibitem{HAMA}
K. Hamada, Prog. Theor. Phys. 103 (2000) 1237, Prog. Theor. Phys. 105
       (2001) 673.
\end{thebibliography}
\end{document}